\documentclass[twocolumn, aps, prl]{revtex4}

\usepackage{amssymb,amsmath}
\usepackage{graphicx, epsfig}

\def \hf{\tfrac{1}{2}}    

\def \ord{\mathcal{O}}

\def\lbc{\left[}    \def\rbc{\right]}

\newcommand{\ket}[1]{\left|{#1}\right.\rangle}
\newcommand{\xpct}[1]{\langle{#1}\rangle}    

\newcommand{\nup}{N_{\uparrow}}

\begin{document}

\title{Edge-locking and quantum control in highly polarized spin chains}

\author{Masudul Haque}
\affiliation{Max-Planck Institute for the Physics of Complex
Systems, N\"othnitzer Str.~38, 01187 Dresden, Germany}

\begin{abstract}

For an open-boundary spin chain with anisotropic Heisenberg (XXZ)
interactions, we present states in which a connected block near the edge is
polarized oppositely to the rest of the chain.  We show that such blocks can
be `locked' to the edge of the spin chain, and that there is a hierarchy of
edge-locking effects at various orders of the anisotropy.  The phenomenon
enables dramatic control of quantum state transmission: the locked block can
be freed by flipping a single spin or a few spins.

\end{abstract}




\maketitle

\emph{Introduction}  ---
Quantum state transfer, and more generally the control of quantum states, has
in recent years entered the realm of experimental possibilities due to rapid
advances in nanostructure and cold-atom technologies, and as a result has been
the focus of intense theory interest \cite{Bose_PRL03, Bose_ContemPhys07,
ChristandlDattaEkertLandahl_PRL04, RomitoFazioBruder_PRB05,
DeChiaraRossiniMontangeroFazio_PRA05, RossiniGiovannettiMontangero_NJP08,
Romero-IsartEckertSanpera_PRA07,
KarbachStolze_PRA05, LyakhovBruder_NJP05}.
More generally, explicit temporal evolution of many-body quantum states far
from equilibrium are no longer of academic interest only, as was the case in
the traditional bulk solid-state context.

In this work, we present and analyze a phenomenon associated with the
high-energy spectrum of open-boundary spin chains, namely, the locking of spin
states by the edge.  We also show how this `edge-locking' effect can be
exploited to exert control over state propagation and spin transfer processes
in spin chains.


We will consider spin-$\frac{1}{2}$ chains governed by an anisotropic
Heisenberg interaction, \emph{i.e.}, $XXZ$ chains.  We consider highly
polarized states, and show that a block of spins anti-aligned to this
background can have stable positions if placed appropriately at or near the
edge.  We reveal the sense in which these configurations are close to being
stationary states of the Hamiltonian.  The presence of such stable
arrangements, and the possibility to convert to non-stable formations via
operations on a few spins, open up simple but powerful possibilities for
controlling spin state propagation.


The $XXZ$ chain is a basic model of condensed matter physics, and has long
been the subject of sustained theoretical activity.  The open chain has
received far less detailed attention than the periodic case, and even less
material is available for physics far from the ground state.
%
%
A new localization phenomenon associated with the $XXZ$ chain is thus
obviously of fundamental interest.  In addition, the $XXZ$ model has recently
been shown to describe Josephson junction arrays of the flux qubit type
\cite{LyakhovBruder_NJP05}, and may also be realizable in optical lattices
\cite{DuanDemlerLukin_PRL03}.  The mechanisms for quantum control uncovered by
our results should be possible to implement in one of these setups in the
foreseeable future.

\emph{Hamiltonian} ---
The open antiferromagnetic $XXZ$ chain with $L$ sites is described by the
Hamiltonian
\[
H_{XXZ} ~=~ J_x \sum_{j=1}^{L-1}\lbc S_j^xS_{j+1}^x + S_j^yS_{j+1}^y ~+~
{\Delta}S_j^zS_{j+1}^z \rbc 
\; .
\]
The $S^zS^z$ term acts as an `interaction' penalizing alignment of neighboring
spins.
The in-plane terms $(S_j^xS_{j+1}^x+S_j^yS_{j+1}^y) = \hf(S_j^+S_{j+1}^- +
S_j^-S_{j+1}^+)$ provide `hopping' processes relevant for quantum state
transfer.  Since $H_{XXZ}$ preserves total $S^z$, the dynamics is always
confined to sectors of fixed numbers $\nup$ of up-spins.
We will mostly consider the large $\Delta$ regime, where the localization
phenomena to be described are most robust.  Energy and time are measured in
$J_z= J_x\Delta$ and $J_z^{-1} = (J_x\Delta)^{-1}$ units unless stated
otherwise.

\begin{figure}
\centering
 \includegraphics*[width=0.95\columnwidth]{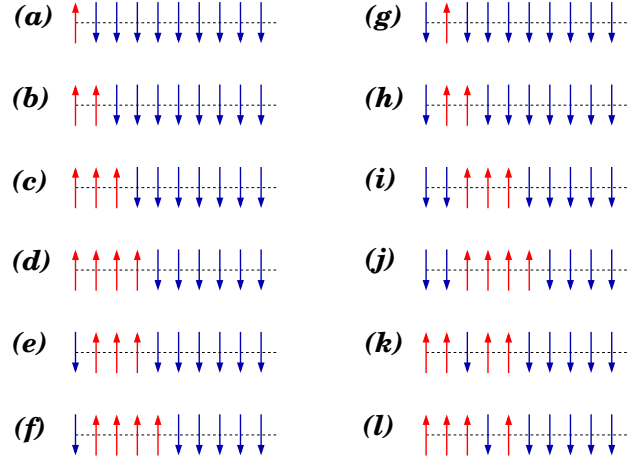}
\caption{ \label{fig_configurn-cartoons}
(Color online.)  A few $\uparrow$ spins at the left edge of an almost
  polarized spin chain. The leftmost 10 spins are shown; the remaining spins
  to the right are all $\downarrow$'s.
Configurations \emph{a}--\emph{f} are edge-locked, while \emph{g}--\emph{l} are
not.
%
%
}
\end{figure}

\emph{Edge-locked states} --- In Fig.\ \ref{fig_configurn-cartoons} we show
some example configurations, \emph{i.e.}, positions of spin-up blocks near the
edge, in a background of down spins.  In the configurations shown on the left,
the $\uparrow$ blocks are locked by the edge at large $\Delta$, while the
blocks in the right-column configurations are not edge-locked.  In dynamic
terms, this means that the configurations on the left column are
\emph{stable}, while those shown on the right decay away.  Of course,
`stability' should be understood in terms of timescales relevant to the
edge-locking physics, and are not absolute.

The simplest and most robust edge states are those in which the block starts
at the very edge site, such as configurations \emph{a} through \emph{d} in
Fig.\ \ref{fig_configurn-cartoons}.  Even a single $\uparrow$ spin placed in
this way is localized at the edge.  For $\nup$ up-spins placed this way, we
call these configurations $\ket{L_{\nup,(1)}}$ or $\ket{R_{\nup,(1)}}$,
depending on whether the block is at the left edge or right edge of the chain.
Figs.\ \ref{fig_configurn-cartoons}\emph{a} through
\ref{fig_configurn-cartoons}\emph{d} are thus $\ket{L_{1,(1)}}$ through
$\ket{L_{4,(1)}}$.
The subscript `(1)' indicates that the $\uparrow$-blocks start at the very
edge site.

The second class of edge states are those where the block starts at the $j=2$
site (or ends at $j=L-1$).  We call these $\ket{L_{\nup,(2)}}$ or
$\ket{R_{\nup,(2)}}$.  For such states to be edge-locked, one needs blocks of
three or more $\uparrow$'s, \emph{i.e.}, $\nup\geq3$.  This is indicated in
Fig.\ \ref{fig_configurn-cartoons} by showing $\ket{L_{1,(2)}}$,
$\ket{L_{2,(2)}}$ on the right column (not edge-locked) as \emph{g}, \emph{h},
and $\ket{L_{3,(2)}}$, $\ket{L_{4,(2)}}$ on the left column (edge-locked) as
\emph{e}, \emph{f}.
Similarly, blocks starting at $j=3$ (or ending at $j=L-2$) are stable only for
$\nup\geq5$.  Generalizing, state $\ket{L_{\nup,(k)}}$ having a block starting
at site $k$ will be stable via edge-locking only if $\nup{\geq}(2k-1)$.

The edge-locking effects are due to spectral separation of the stable states
from other states, which prevents hybridizations that might enable propagation
of the oppositely-polarized blocks.  The spectral separation can be
understood using perturbative arguments at small $\Delta^{-1}$.  In the
hierarchy described above, the first class of edge-locking (blocks starting at
the edge site) is a zeroth-order effect, while edge-locking at the second
level (blocks starting at next-to-edge site) is an $\ord(\Delta^{-2})$ effect.
Generally, the level-$k$ edge-locking of this hierarchy is an
$\ord(\Delta^{-2(k-1)})$ effect.

\begin{figure}
\centering
 \includegraphics*[width=0.95\columnwidth]{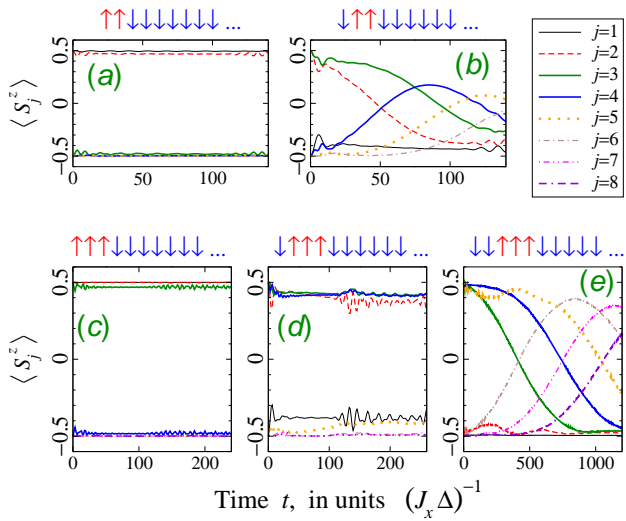}
\caption{ \label{fig_evolution-1}
(Color online.)  Dynamics of 18-site open $XXZ$ chain ($\Delta=4$), initiated
  with $\nup=2$ or $\nup=3$ oppositely-polarized spins near the left edge.
  Initial configurations are shown on top of each panel.  In the edge-locked
  cases, local spin values $\xpct{S_j^z}$ do not vary much from their initial
  values.
%
%
}
\end{figure}

\emph{Temporal dynamics} --- 
Fig.\ \ref{fig_evolution-1} demonstrates the edge-locking phenomenon through
explicit time evolution of several configurations.  
The top row shows the evolution of $\nup=2$ states $\ket{L_{2,(1)}}$ and
$\ket{L_{2,(2)}}$.  The first is an edge-locked state and shows very little
evolution, while the second is not locked, and thus the $\uparrow\uparrow$
block propagates to the right.

The bottom row shows $\nup=3$ states.  Now there are two states where the
$\uparrow\uparrow\uparrow$ block is locked by the left edge.  We have chosen a
moderate value of $\Delta$ so that the $\ord(\Delta^{-2})$ locked state
$\ket{L_{3,(2)}}$ can be clearly seen to have weaker locking than the
$\ord(\Delta^0)$ locked state $\ket{L_{3,(1)}}$.  (Fig.\
\ref{fig_evolution-1}\emph{d} has more dynamics and larger oscillations than
\ref{fig_evolution-1}\emph{c}.)  Obviously, the higher-order locking can be
made more robust by using a larger $\Delta$.

The numerical results of Fig.\ \ref{fig_evolution-1} are for 18-site open
chains, but a longer chain would display identical time evolutions at the time
scales shown.  The size plays a role only when the propagating block meets the
other edge and gets reflected.  It is clear that the unlocked blocks in Fig.\
\ref{fig_evolution-1} (\emph{b},\emph{e}) are still propagating to the right
at the timescales shown.

\begin{figure}
\centering
 \includegraphics*[width=0.95\columnwidth]{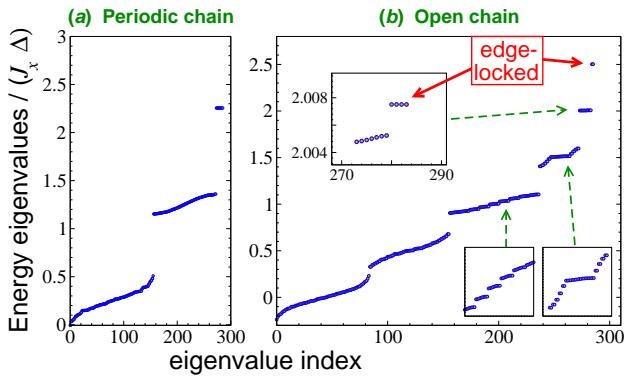}
\caption{ \label{fig_spectra-1}
(Color online.) Energy spectra: $L=13$, $\nup=3$, $\Delta=10$.  Compared to
  the periodic chain, the open chain spectrum has extra features, some of
  which are highlighted in insets.  With $\nup=3$, there are two classes of
  edge-locked states, the two states of the top band (main plot) and two of
  the four states separating out from the next band (upper inset).
}
\end{figure}

\emph{Spectral explanation} --- We now turn to the spectral separation that
causes edge-locking by suppressing hybridization with non-locked states.
Fig.\ \ref{fig_spectra-1} shows the energy spectrum for $L=13$ sites, in the
$\nup=3$ sector.
At large $\Delta$, the spectrum separates into well-separated `bands'.  In the
periodic chain, the bands correspond to cases where the three $\uparrow$ spins
are next to each other (top band), or two are next to each other (middle
band), or no two $\uparrow$ spins neighbor each other (bottom band).  In
general, with $\nup$ ($<L/2$) up spins in a periodic large-$\Delta$ chain, the
spectrum is separated into $p(\nup)$ bands, where $p(n)$ is the number of
integer partitions of $n$.  The topmost band is maximally ferromagnetic, and
has the minimal number (two) of favorable $\uparrow$-$\downarrow$ bonds and
($L-2$) non-favorable ($\uparrow$-$\uparrow$ or $\downarrow$-$\downarrow$)
bonds.

Fig.\ \ref{fig_spectra-1}\emph{b} shows the effect of open boundaries.  The
spectrum described above for the periodic chain now acquires an explosion of
additional features.  The periodic-chain bands get split, because the edge
allows additional possibilities for numbers of favorable and unfavorable
bonds.  In addition, several of these new bands have additional
sub-structures, as shown in insets.  While these structures are all
interesting,  
in this work we will only be concerned with the top two bands of the open
chain, which both emerge from the topmost band of the periodic chain, and
hence are related to periodic-chain configurations with all $\uparrow$ spins
in a connected block.

The top-most band has only two states, and these are the most obvious
edge-locked states.  For three up spins at $\Delta^{-1}=0$, this is a
degenerate two-dimensional manifold spanned by $\ket{L_{3,(1)}} =
\ket{\uparrow\uparrow\uparrow\downarrow\downarrow\downarrow\downarrow\downarrow...}$
and $\ket{R_{3,(1)}} =
\ket{...\downarrow\downarrow\downarrow\downarrow\downarrow\uparrow\uparrow\uparrow}$.
At finite $\Delta^{-1}$, other configurations contribute to the two states,
but for small enough $\Delta^{-1}$ the eigenstates are dominated by
$\ket{L_{3,(1)}}\pm\ket{R_{3,(1)}}$.  Similarly, for other values of
$\nup$, the topmost eigenstates are dominated by
$\ket{L_{\nup,(1)}}\pm\ket{R_{\nup,(1)}}$ at large $\Delta$.
These two states appear at the very top because $\ket{L_{\nup,(1)}}$ and
$\ket{R_{\nup,(1)}}$ are the only configurations having a \emph{single}
favorable anti-aligned bond and thus a maximum number ($L-2$) of unfavorable
bonds.  In contrast, in the periodic case any configuration has at least two
favorable bonds.

The edge block in $\ket{L_{\nup,(1)}}$ is strongly locked because this state
is hybridized mainly with $\ket{R_{\nup,(1)}}$.  From $\ket{L_{\nup,(1)}}$, it
is energetically possible to tunnel into the $\ket{R_{\nup,(1)}}$ state, but
such a process is exponentially suppressed at large chain lengths.  Thus
$\ket{L_{\nup,(1)}}$ can be regarded as `stationary' for practical purposes,
as Fig.\ \ref{fig_evolution-1}(\emph{a},\emph{c}) demonstrates dynamically.

The other edge-locking effects are weaker and can be seen by zooming into the
second band from the top, which consists of configurations with two favorable
bonds (Fig.\ \ref{fig_spectra-1}\emph{b} upper inset).  Four states separate
out from the rest of this band, with $\ord(\Delta^{-2})$ splitting.  These
four states are linear combinations of $\ket{L_{3,(2)}}$, $\ket{R_{3,(2)}}$,
and
\[
\ket{\uparrow\uparrow\downarrow\downarrow\downarrow.....\downarrow\downarrow\downarrow\uparrow}
\quad {\rm and} \quad
\ket{\uparrow\downarrow\downarrow\downarrow.....\downarrow\downarrow\downarrow\uparrow\uparrow}
\; .
\]
The rest of the band is dominated by linear combinations of the remaining
configurations containing the $\uparrow$ spins in connected blocks farther
from the edge.

Due to the spectral separation, the four states are not hybridized with the
remaining block configurations.  This locks the $\ket{L_{3,(2)}}$ and
$\ket{R_{3,(2)}}$ configurations to their respective edges, because from any
of these states, tunneling to the other three of the sub-manifold is a very
high-order process in $J_x$.

\begin{figure}
\centering
 \includegraphics*[width=0.95\columnwidth]{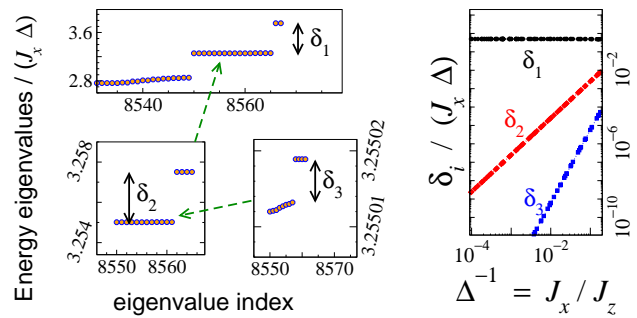}
\caption{ \label{fig_spectra-2}
(Color online.)  Left: hierarchy of sub-band splittings seen by zooming in
  successively.  Here  $\nup=5$ ($L=18$, $\Delta=10$), so the
  first three levels of the hierarchy are present.
Right: Energy splittings scale as $\delta_1\sim\Delta^0$,
  $\delta_2\sim\Delta^{-2}$, $\delta_3\sim\Delta^{-4}$.
}
\end{figure}

\emph{Hierarchy in spectrum} ---
For $\nup=3$, only the first two classes of edge-locked states are present, as
indicated by the two solid arrows in Fig.\ \ref{fig_spectra-1}\emph{b}.  An
additional level of the hierarchy becomes available with each increase of
$\nup$ by two.  The associated spectral separations can be seen by
successively zooming in, within the next-to-top band.  Fig.\
\ref{fig_spectra-2} (left three panels) shows this for $\nup=5$, where three
edge-locked configurations appear.  Fig.\ \ref{fig_spectra-2} (right)
displays the associated gaps scaling as $\delta_k\sim\Delta^{2k-1}$.

\emph{Physical reason for spectral separation} ---
The spectral separations leading to edge-locking can be understood through
degenerate perturbation theory.  We give a brief explanation for the second
level, namely the separation of states $\ket{L_{\nup,(2)}}$ and
$\ket{R_{\nup,(2)}}$ from the states $\ket{L_{\nup,(k)}}$ with $2<k<(L-1)$.

At $\Delta^{-1}=0$, the configurations with two favorable bonds are all
degenerate.  At small finite $\Delta^{-1}$ these spread out to form the
next-to-top band.  The hybridization of these levels happens at order
$\Delta^{-\nup}$, because $\nup$ `hoppings' are required to connect
configurations $\ket{L_{\nup,(k)}}$ and $\ket{L_{\nup,(k+1)}}$.

On the other hand, since each $\ket{L_{\nup,(k)}}$ is connected to itself by
two hoppings, the states acquire energy shifts at order $\Delta^{-2}$.
Considering the energies of the intermediate states in this process, one can
see that the states $\ket{L_{\nup,(2)}}$ and $\ket{R_{\nup,(2)}}$ have a
different energy shift compared to the rest, due to the edge.  For $\nup>2$
this $\ord(\Delta^{-2})$ effect is stronger than the $\ord(\Delta^{-\nup})$
hybridization, so that these states separate out without hybridizing with the
rest, and are thus edge-locked.

The argument can be extended to higher stages of the hierarchy.  Blocks
starting at site $k$ have a $\ord(\Delta^{-2(k-1)})$ shift distinct from the
shift of the farther blocks, and hence will be separated from the band if
$2(k-1)<\nup$, allowing further edge-locked configurations.

\emph{State transfer protocols} ---
The edge-locking phenomenon provides many opportunities for controlling the
evolution and transport of spin states, provided that the experimental
realization of the $XXZ$ chain allows single-site (or few-site) addressing.
We point out a few of the most obvious quantum control protocols.

If single-site spin flipping probes ($\pi$ pulse) can be implemented, this can
be used to `release' a locked block.  For example, by flipping the first spin
of the locked block in $\ket{L_{3,(1)}}$, one gets the state
$\ket{L_{2,(2)}}$, in which the two-site block is not locked (Fig.\
\ref{fig_evolution-1}\emph{b}), and so starts propagating.  Similarly,
starting with a 5- or 6-site oppositely-polarized block at the edge, applying
a $\pi$ pulse on the first \emph{two} sites initiates the transmission of a
signal consisting of a block of spins anti-aligned to the background.  Once
the signal reaches the other edge of the open chain, the block could also be
locked to the other edge by $\pi$-pulsing one or two spins at the other edge
at the appropriate moment.

More complex dynamics can be launched by applying a $\pi$ pulse to a site
internal to the locked block, for example, by flipping the second site of the
locked $\ket{L_{4,(1)}}$ configuration.  The resulting
$\uparrow\downarrow\uparrow\uparrow\downarrow\downarrow\downarrow\downarrow....$
has the following dynamics: the $\uparrow$ at site 3 moves to site 2 so that a
two-site $\uparrow\uparrow$ block then stays locked to the edge, while a third
$\uparrow$ propagates to the right.  While a complete explanation involves the
lower energy bands, which are beyond the scope of this paper, the tendency to
lock blocks at the edge is clearly seen here too.

The hierarchy structure can be also used to design more subtle control protocols,
for example, a $\pi$ pulse can be used to go from a strongly locked state to a
weakly locked state; the difference is especially acute at moderate
$\Delta\gtrsim1$.  For example, starting with  $\ket{L_{4,(1)}}$ and flipping
the first spin performs such an operation.


\emph{Spinless fermion model} ---
The $XXZ$ chain model is generally considered to be equivalent to the spinless
fermion model with nearest-neighbor couplings.  However, if the interaction is
of $Vn_in_{i+1}$ form ($n_i$ are site occupancies), the spectral structure
associated with open-chain edge-localization is quite different from the $XXZ$
chain, as can be seen by comparing present results with
Ref.~\cite{PintoHaqueFlach_PRA09}.  The physics becomes identical if one uses
the $V(n_i-\hf)(n_{i+1}-\hf)$ form, which involves interactions between
unoccupied sites.

\emph{Experimental realizations} --- 
The most obvious realizations are bulk materials with chain structures.  There
are several compounds whose spin physics are reasonably well-described by
$\Delta>1$ $XXZ$ Hamiltonians, such as CsCoCl$_3$ 
with $\Delta\sim7$ 
\cite{YoshizawaHirakawaSatijaShirane_PRB81, GoffTennantNagler_PRB95}, and
BaCo$_2$V$_2$O$_8$ with $\Delta\sim2$ \cite{KimuraOkunishi-etal_PRL07}.
Unfortunately, single-site addressing is generally not feasible, and
nonequilibrium states generally relax rapidly to the ground state in bulk
materials.  Nevertheless, it may still be possible to probe the physics
presented in this paper.  With the spins polarized completely by a magnetic
field above the saturation threshold, a localized excitation (through neutrons
or laser pulse) could depolarize a few sites, moving the system to a sector
where edge states can be relevant.  Applying an excitation near one end of the
material, one can watch for response at the other end, which would indicate
whether the excited block is locked or propagating.


A more promising route is through Josephson junction arrays, which in some
arrangements (persistent-current qubits
\cite{MooijOrlandoLevitov-etal_Science99, LevitovOrlandoMooij_arxiv01}) are
well-described by an $XXZ$ Hamiltonian \cite{LyakhovBruder_NJP05}.
Such arrays could be prepared in non-equilibrium initial states and
single-qubit addressing should be straightforward, and thus might be ideal for
initial exploration of the physics described here.

Finally, there is the possibility of realizing $XXZ$ lattices in cold-atom
setups \cite{DuanDemlerLukin_PRL03}, although this has additional
complications of realizing a well-defined edge and accounting for harmonic
traps.


\emph{Open issues} --- 
This work raises several issues demanding further investigation.  One expects
dynamical effects and additional localization phenomena associated with the
sub-structures of the lower bands (Fig.\ \ref{fig_spectra-1}), which are yet
to be explored.  Depending on the experimental realization(s) that become
available, the effects of terms beyond the $XXZ$ Hamiltonian relevant for the
particular realization need to be analyzed, and new control protocols can be
designed for the site addressing methods that are possible.  Finally, the
$XXZ$ model being Bethe-ansatz solvable even with open boundary conditions
\cite{Alcaraz-et-al_JPhysA87}, it remains an open problem to find out how the
sub-band and edge-locking structures in the high-energy spectrum are reflected
in the Bethe ansatz root structure.

%

%
%
The author thanks J.-S.~Caux, S.~Flach, A.~M.~L\"auchli, and T.~Vojta for
helpful discussions.


\end{document}